\documentclass[journal]{IEEEtran}
\usepackage{cite}

\ifCLASSINFOpdf
\else
  \usepackage[dvips]{graphicx}
  \DeclareGraphicsExtensions{.eps}
\fi
\usepackage[cmex10]{amsmath}
\usepackage[tight,footnotesize]{subfigure}

\usepackage{fixltx2e}

\usepackage{stfloats}
\begin{document}
%
\title{Ultra-low vibration pulse-tube cryocooler stabilized cryogenic sapphire oscillator with $10^{-16}$ fractional frequency stability}
%
%
%

\author{John~G.~Hartnett and ~Nitin~R.~Nand,
\thanks{John~G.~Hartnett and ~Nitin~R.~Nand are with the School of Physics, the University of Western Australia, Crawley, 6009, W.A., Australia.}
\thanks{Manuscript received March 20, 2010; ....}}

%
%

\markboth{IEEE Trans. on Microwave Theory and Techniques,~Vol.~10, No.~1, December~2010}%
{Shell \MakeLowercase{\textit{et al.}}: Bare Demo of IEEEtran.cls for Journals}
%



\maketitle

\begin{abstract}
A low maintenance long-term operational cryogenic sapphire oscillator has been implemented at 11.2 GHz using an ultra-low-vibration cryostat and pulse-tube cryocooler. It is currently the world's most stable microwave oscillator employing a cryocooler. Its performance is explained in terms of temperature and frequency stability. The phase noise  and the Allan deviation of frequency fluctuations have been evaluated by comparing it to an ultra-stable liquid-helium cooled cryogenic sapphire oscillator in the same laboratory. Assuming both contribute equally, the Allan deviation evaluated for the cryocooled oscillator  is $\sigma_y \approx 1 \times  10^{-15}\tau^{-1/2}$ for integration times $1 < \tau < 10$ s with a minimum $\sigma_y = 3.9 \times 10^{-16}$ at $\tau = 20$ s. 
The long term frequency drift is less than $5 \times 10^{-14}$/day.
From the measured power spectral density of phase fluctuations the single side band phase noise can be represented by $\cal{L}$$_{\phi}(f) = 10^{-14.0}/f^4+10^{-11.6}/f^3+10^{-10.0}/f^2+10^{-10.2}/f+ 10^{-11.0} \, rad^2/Hz$ for Fourier frequencies $10^{-3}<f<10^3$ Hz in the single oscillator. As a result $\cal{L}$$_{\phi} \approx -97.5 \; dBc/Hz$ at 1 Hz offset from the carrier. 
\end{abstract}

\begin{IEEEkeywords}
cryogenic sapphire oscillator, cryocooler, phase noise, frequency stability.
\end{IEEEkeywords}

%
\IEEEpeerreviewmaketitle

\section{Introduction}
%
%
%
%
\IEEEPARstart{H}{igh} purity single crystal sapphire has extremely low loss at microwave frequencies \cite{Braginsky} and at cryogenic temperatures and as a result it has been used to develop the most highly stable microwave oscillators on the planet \cite{Hartnett2006, Hartnett2010}.  Loop oscillators are operated with a cryogenic sapphire resonator tuned to a high order whispering-gallery mode. This acts both as a very narrow loop filter and a ultra-high Q-factor frequency determining element in the Pound servo used to lock the oscillator frequency to the very narrow natural resonance line in the sapphire \cite{Locke2008}.  

Cryogenic sapphire oscillators have been deployed to a number of frequency standards labs \cite{Watabe2006, Watabe2007} and have facilitated atomic fountain clocks  reaching a performance limited only by quantum projection noise \cite{Santarelli1999}.  Since most atomic fountain clocks \cite{Wynands2005} have significant dead time in their interrogation cycle, simply due to the fact that they are pulsed devices where one has to wait until the detection process is finished before the next cloud of cold atoms is launched, it is impossible to avoid influences of the Dick effect \cite{Dick1987, Dick1990, Audion1998, Santarelli1998}. The Dick effect is the introduction of the phase noise of the local reference oscillator, from which the microwave cavity frequency is derived, to the short term stability of the fountain clock. One way  to reduce this is to use a lower noise highly stable reference, and this is where the cryogenic sapphire oscillator has been used.

Recently a new project was started at the University of Western Australia to replicate the state-of-the-art cryogenic sapphire oscillator but using an ultra-low vibration designed cryostat \cite{Chao} and a pulse tube cryo-cooler \cite{Hartnett2010}.  The latter paper reported only on the frequency stability when using the low vibration cryocooler and cryostat under different operating conditions. Improvements have since been made and the design optimized. This paper details the cryogenic sapphire oscillator performance after those changes. For the first time we report on the resonator temperature stability,  on the oscillator phase noise  over 6 decades of offset Fourier frequencies from the carrier and also on the oscillator stability evolution for integration times $0.1 \, s < \tau < 10,000 \, s$) and what are the limiting noise processes with a view to future improvements.

Previously cryocoolers have been used to cool the sapphire element in cryogenic sapphire oscillators. One was used in oscillators developed for NASA's deep space tracking network \cite{Dick2, Wang}, which were built using a Gifford-McMahon cycle cryocooler, but still had limited performance at about $10^{-14}$ over 1 s of averaging and a fractional frequency drift of about $10^{-13}$/day  \cite{Wang2}.  Later, Watabe et al. \cite{Watabe2003} used a two-stage pulse-tube cryocooler with a lower level of vibration than the Gifford-McMahon type, but the stability was still above $10^{-14}$.  The pulse-tube design described in this paper is much lower in vibration at the cryocooler head and incorporates more vibration reduction features \cite{Chao} to effectively reduce vibrations to a point where they do not impact the operation of the oscillator \cite{Hartnett2010}. Similar efforts have been made at the FEMTO-ST Institute, Besan\c{c}on, France. There, a cryogenic oscillator was built for ESA's deep space tracking network and it has a fractional frequency stability of about $2 \times 10^{-15}$ from 1 to 1000 s of averaging \cite{Grop1, Grop2}.

This current project has been to develop a local oscillator for VLBI radio astronomy, where an improved short-term stability of about two orders of magnitude over the hydrogen maser offers significant gains. Due to the high altitude of very high frequency VLBI sites it may be that the quality of the receiver signal is limited by the local oscillator. If that is the case, by averaging over shorter time intervals using a cryogenic sapphire oscillator as the local oscillator will result in clearer images.  

\section{Ultra-low vibration cryostat}
\begin{figure}[!t]
\centering
\includegraphics[width=3.5in]{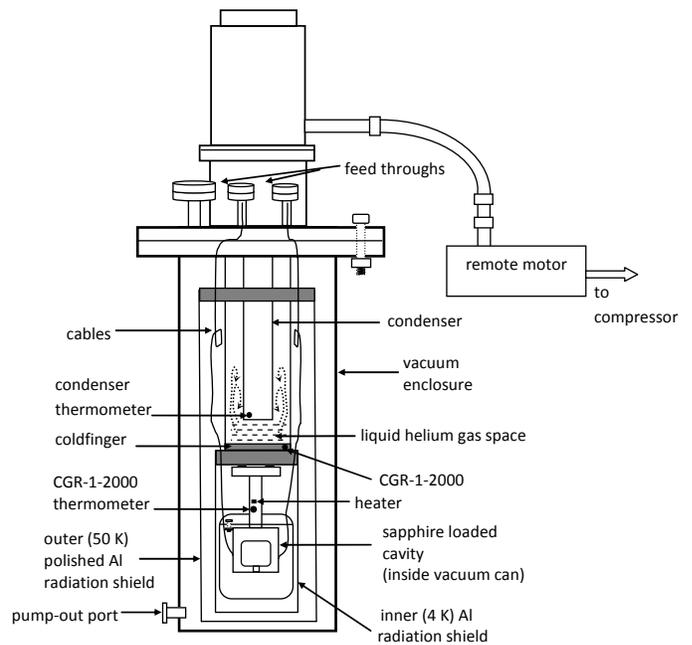}
\caption{Schematic of cryocooler cryostat showing vacuum enclosure and the location of sapphire loaded cavity on the ``coldfinger'' that is cooled by a few liters of liquid helium maintained by the condenser inside the helium gas space. The sapphire temperature is actively controlled using a carbon glass CGR-1-2000 sensor, a power resistor as a heater and a Lakeshore 340 temperature controller. The sensor and the heater are  located on the ``copper post'' supporting the sapphire resonator. }
\label{fig_1}
\end{figure}

In order to reach the stability levels achieved in liquid helium cooled sapphire oscillators the vibrations produced by any cryocooler need to be reduced or compensated for in some way, so that they have minimal effect on the frequency of the oscillator determined by the cryogenic resonator. Wang et al. \cite{Wang2} used a jacket of helium gas to disconnect the motion of the condenser from the cryogenic resonator. The same approach has been adopted here \cite{Chao}. See Fig. \ref{fig_1}. 

In this case a helium gas space was designed in the first version of the cryostat primarily to allow for operation with positive pressure in that space. Hence a bellows was introduced to free the cold head and hence the ``condenser'' from any rigid links to the resonator.  At the condenser there is approximately a 12 $\mu m$ vertical motion at about 1.4 Hz. The resonator inside its own vacuum can is attached to the base of the space where the condenser maintains a few liters of liquid helium \cite{Chao}. This is labeled ``coldfinger'' in Fig. \ref{fig_1}. 

It was found however that there was a strong effect on the frequency stability of the oscillator at about 1000 s of averaging when run with a pressure near one atmosphere \cite{Hartnett2010}. The best condition was found where the lowest pressure (about 50 kPa in practice) was allowed to develop in the helium gas space. Still there is sufficient gas there to provide the convective link to recool the helium as it evaporates. The temperature was maintained at about 3.35 K in the liquid  under these conditions. The short term frequency stability (measured at an averaging time of $\tau$ = 1 s) was found to slightly worse but the long term stability was greatly improved. Therefore the bellows was unnecessary and it was removed in a redesign of the cryostat. 

\begin{figure}[!t]
\centering
\includegraphics[width=3.5in]{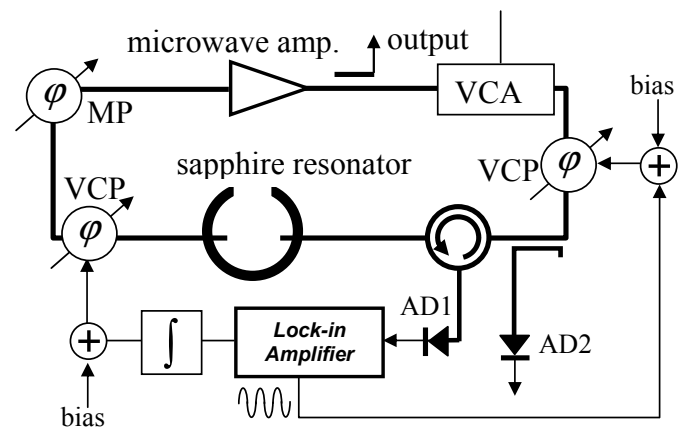}
\caption{Schematic of microwave loop oscillator. The symbols are as follows: MP is the mechanical phase-shifter to achieve the correct loop phase for sustained oscillation, VCPs are voltage controller phase-shifters used for modulation and error correction of the loop phase lengths, and VCA is the voltage controlled attenuator used to servo control the circulating power in the loop by comparing the output of AD2 (an amplitude detector) to a voltage reference (circuit not shown). AD1 is the amplitude detector connected to the resonator reflection port, the signal from which is used as the input to the Lock-in amplifier to generate the frequency control error signal to servo control the microwave oscillator frequency to that of the resonance.}
\label{fig_2}
\end{figure}
\begin{figure}[!t]
\centering
\includegraphics[width=3.5in]{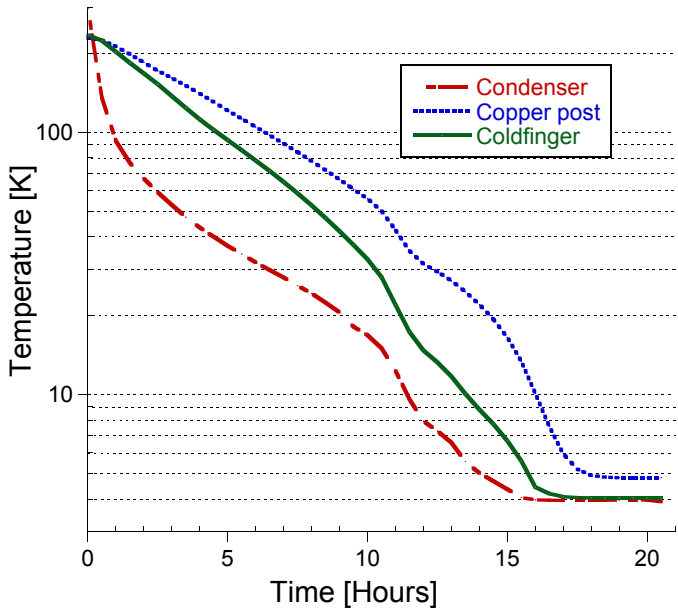}
\caption{(color online) Temperatures at the three points monitored: the condenser, the coldfinger and on the copper post supporting the cavity. The curves indicate the cool-down time from initial start of cryocooler compressor. The sensors on the coldfinger and the copper post are uncalibrated above 100 K.}
\label{fig_3}
\end{figure}
\begin{figure}[!t]
\centering
\includegraphics[width=3.5in]{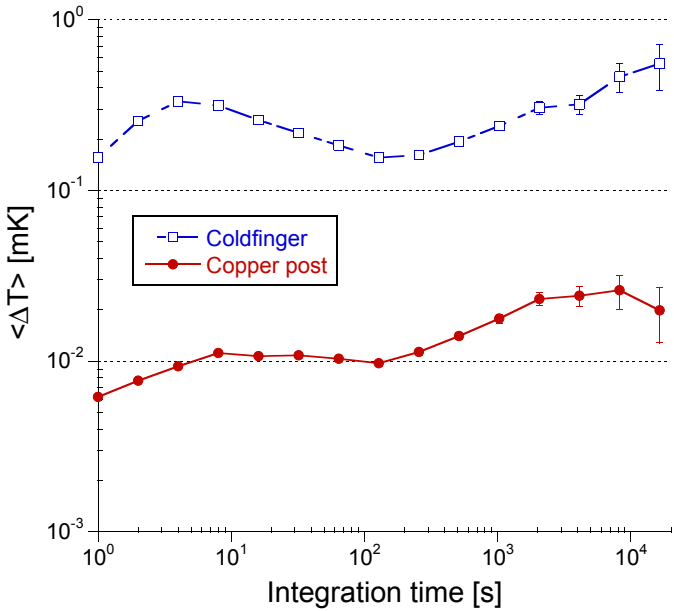}
\caption{(color online) $\left\langle \Delta T \right\rangle$ the \textit{rms} temperature fluctuations expressed in units of [mK], calculated from the Allan deviation algorithm applied to time sampled (with a gate time of 1 s) temperatures at both the coldfinger and the copper post.  }
\label{fig_4}
\end{figure}

\section{Cryogenic Sapphire Oscillator}
A highly stable oscillator employing a cryogenic temperature stabilized Crystal Systems HEMEX grade \cite{CS} single-crystal sapphire resonator, cooled with a 2-stage CryoMech PT407-RM ultra-low-vibration pulse-tube cryocooler, has been constructed. The new cryocooled sapphire oscillator contains a nominally identical resonator to those in the liquid helium cooled cryogenic sapphire oscillators used in the lab. The design of sapphire resonators used in these cryogenic oscillators has been previously discussed \cite{Tobar2006} as has the design of the loop oscillator and its servo control systems \cite{Hartnett2006, Locke2008}. See Fig. 2. However in this case we derive 17 dBm of output power from the microwave amplifier in the loop.

The chosen operational mode  is the whispering-gallery $WGH_{16,0,0}$ mode with a frequency of 11.202 GHz. The mode nomenclature means it has 16 azimuthal variations in the electromagnetic field standing wave around half its circumference, and one variation each along the radial and axial axes.  The new resonator exhibited a turning point in its frequency-temperature dependence at about 5.9385 K and has a loaded Q-factor of $1.05 \times 10^9$ at the turnover temperature. The turnover temperature was previously reported \cite{Hartnett2010} as 5.984 K but this difference was found to be due to thermal gradients, which has been reduced by the introduction of a 4 K radiation shield, made from thin (0.5 mm thick) aluminum (Fig. \ref{fig_1}).  

The resonator primary and secondary port coupling coefficients at the turnover temperature were set to 0.80 and about 0.01, respectively. The microwave energy is coupled into the resonator with loop antenna probes, through the lateral cavity wall, made from the same coaxial cables that connect it to the loop oscillator in the room temperature environment. Using an Endwave JCA812-5001 amplifier with sufficient microwave gain (nearly 50 dB), a microwave filter set at the resonance frequency and the correct loop phase, set via a mechanical phase shifter, we get sustained oscillation.  The details of the loop oscillator, the temperature control of the sapphire loaded cavity, the power control of the loop oscillator and the Pound frequency servo locking the oscillator frequency to the resonance in the sapphire crystal are the same as used in the liquid helium cooled oscillators \cite{Locke2008}. 

In the measurements reported here a liquid helium cooled  sapphire oscillator \cite{Hartnett2006}, operating on the same $WGH_{16,0,0}$ mode with a frequency of 11.200 GHz, was used to make frequency and phase comparisons with the new cryocooled sapphire oscillator. 

\section{Temperature Stability}
Carbon glass CGR-1-2000 sensors were thermally anchored to lower vacuum can. One at the base of the coldfinger to monitor the temperature there and another to the copper post supporting the sapphire loaded cavity resonator.  The latter is used to actively control the temperature of the sapphire crystal using a Lakeshore 340 temperature controller. As part of the thermal design we introduced 3 stainless steel washers to thermally isolate the copper post from the coldfinger. By using stainless washers of different thickness one is able to increase the thermal time constant and hence also increase the minimum temperature at which the resonator stabilizes. 

Figure \ref{fig_3} shows the cool down temperatures as a function of time from initially switching on the compressor of the cryocooler. During the cool down slightly more than 1 atmosphere of pressure is maintained in the helium gas space. After 4 K is reached at the coldfinger the valve to the helium gas supply is closed off and a partial vacuum develops in the helium gas space (actually about 50 kPa). The temperature of the liquid helium falls but once the temperature control is activated (to bring the copper post, hence the sapphire, to the control point of 5.9385 K) the coldfinger temperature is raised to and maintained at about 3.35 K. 
\begin{figure}[!t]
\centering
\includegraphics[width=3.5in]{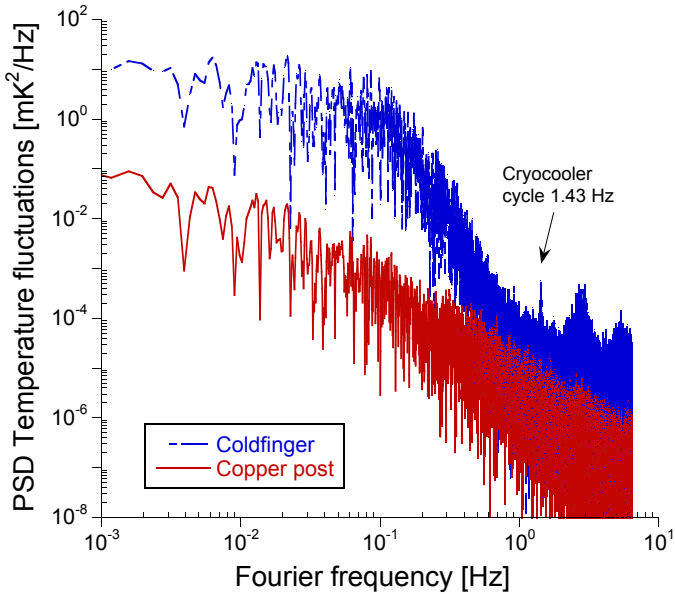}
\caption{(color online) Power spectral density of temperature fluctuations expressed in units of [\mbox{$mK^2$/Hz}] calculated from time sampled  (with a gate time of 0.08 s) temperatures at both the coldfinger and the copper post. The top (blue) spectrum represents the temperature fluctuations at the coldfinger and the bottom (red) spectrum represents the temperature fluctuations at the copper post supporting the sapphire resonator. This is the temperature control point and therefore represents an upper limit to the temperature fluctuations there.}
\label{fig_5}
\end{figure}
\begin{figure}[!t]
\centering
\includegraphics[width=3.5in]{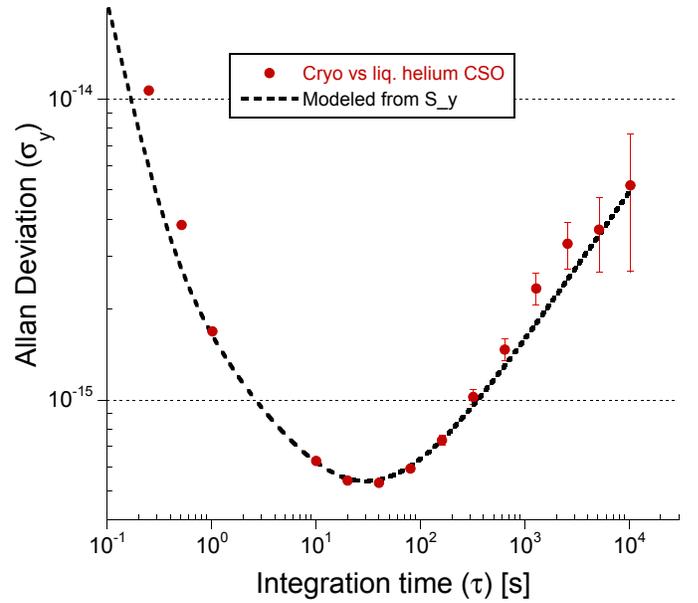}
\caption{(color online) The (red) solid circles represent the Allan deviation $\sigma_y$  calculated from time domain beat data using a $\Lambda$-counter, with gate times of 0.25, 0.5, 1, and 10 s. The (black) dashed line represents the modeled $\sigma_y$ determined from  Eq. (\ref{Sy}) and the single side band phase noise best fit of Eq. (\ref{L}). CSO  = cryogenic sapphire oscillator.}
\label{fig_6}
\end{figure}

The \textit{rms} temperature fluctuations $\left\langle \Delta T \right\rangle$  at the coldfinger and the copper post supporting the sapphire resonator were monitored by taking 150,000 samples with a 1 s gate time of the resistance of the two carbon glass sensors at these points. The data were converted to temperature and $\left\langle \Delta T \right\rangle$  was then calculated from the time series temperature data using the \textit{Stable32} software--the Allan deviation of fractional temperature fluctuations $\sigma_T(\tau)$ was calculated. The latter is related
\begin{equation} \label{sigmaT}
\left\langle \Delta T \right\rangle = \sigma_T(\tau) \; T_0,
\end{equation}
where $T_0$ is the mean temperature being measured. In these measurements there is no mean drift. 

The resulting \textit{rms} temperature fluctuations $\left\langle \Delta T \right\rangle$  at the coldfinger and the copper post supporting the sapphire resonator are shown in Fig. \ref{fig_4}, in units of [mK].  From this it is clear that over the region of most interest, $1<\tau<100$ s, the temperature fluctuations at the copper post supporting the sapphire (hence at the sapphire) are $\left\langle \Delta T \right\rangle \leq 10 \; \mu$K. In fact, since we sampled the control thermometer, in this case, at the `copper post', which was under active servo control, those temperature fluctuations can be only considered an upper limit. In the case of the `coldfinger' data the sensor was only used to monitor the temperature.

At the frequency temperature turnover point ($T_0=5.9385$ K) we calculated the second derivative $$\frac{1}{f_0} \frac{\partial^2 f}{\partial T^2} = 1.98 \times 10^{-9} \; [K^{-2}]$$ by measuring the oscillator beat as we changed the temperature in the cryocooled sapphire oscillator. The latter can be related \cite{Hartnett2002}
\begin{equation} \label{sT}
\sigma_y = \left\langle \frac{\Delta f}{f_0}\right\rangle=\frac{1}{f_0} \frac{\partial^2 f}{\partial T^2} \; \delta T \; \left\langle \Delta T \right\rangle,
\end{equation}
where $f_0$ = 11.202 GHz is the microwave oscillator frequency and $\delta T$ is the error in setting the temperature control set point for the sapphire crystal exactly on the turnover point.

If we solve Eq. (\ref{sT}) for $\delta T$ assuming a value $\left\langle \Delta T \right\rangle$ = 10 $\mu$K, which is the value at $\tau = 10$ s, we get $\delta T$ = 25 mK. Certainly we can find the turnover point much better than that, at least by a factor of 10, so temperature fluctuations do not add any significant noise to the oscillator.

It may be noted also from Fig. \ref{fig_4} that the \textit{rms} temperature fluctuations at the coldfinger are at least an order of magnitude higher than at the copper post. This is due to the thermal filtering of the stainless steel spacers. Note too that at higher frequencies (ie. at shorter times $\tau$) there is a stronger suppression effect. This is particularly indicated in Fig. \ref{fig_5}, where we have converted the  time series temperature data sampled with 0.08 s gate time to power spectral densities of temperature fluctuations.  In the upper spectrum for the coldfinger temperature fluctuations one can clearly see a peak due to the cryocooler compressor cycle frequency of 1.43 Hz. Higher harmonics are apparent too. However in the lower spectrum for the copper post temperature fluctuations one cannot see this; the signal has been significantly attenuated.

\section{Frequency Stability}
The cryocooled oscillator was implemented and time domain data of the 1.58 MHz beat between it and the liquid helium cooled  oscillator measured with an Agilent 53132A $\Lambda$-counter \cite{Dawkins}  with gate times of 0.25, 0.5, 1 and 10 s. From this the Allan deviation ($\sigma_y$) of frequency fluctuations was calculated at integration times $\tau$ equal to the gate times and at even multiples of the gate time with the 10 s gate time data using the \textit{Stable32} software. See Fig. \ref{fig_6} for the results. (When using this type of counter the calculated Allan deviation $\sigma_y$ differs slightly from that calculated from a standard counter. The reader is advised to refer to Ref. \cite{Dawkins} where a full analysis is given.)

\begin{figure}[!t]
\centering
\includegraphics[width=3.5in]{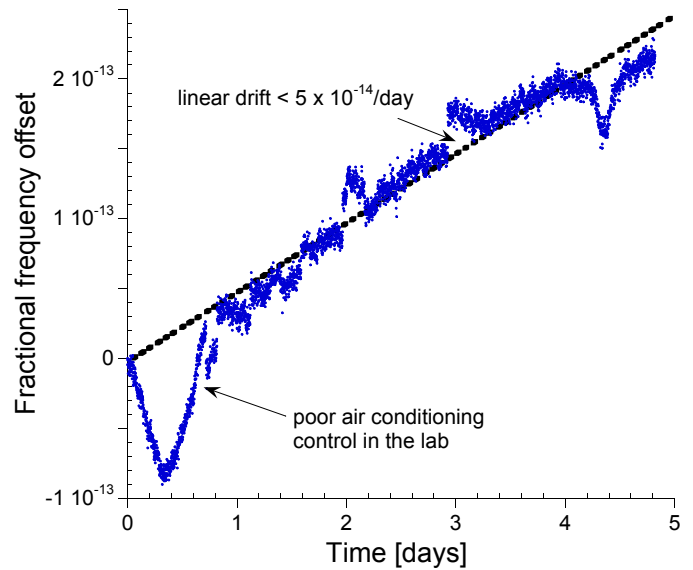}
\caption{(color online) The measured fractional frequency offset of the beat between the cryocooled  sapphire oscillator and an 11.2 GHz signal produced by picking off a high order harmonic of a step recovery diode driven by the doubled 100 MHz signal from our Kvarz hydrogen maser. Each datum is a 100 s gate time sample. The initial large deviation is due to the poor functioning of the air conditioner in the lab, which was later rectified.}
\label{fig_7}
\end{figure}

In the first cryostat design we observed a significant (of order $10^{-12}$/day) linear frequency drift in the oscillator \cite{Hartnett2010}. This has now been significantly reduced to a level of slightly below $5 \times 10^{-14}$/day (see Fig. \ref{fig_7}). It was accomplished largely by the introduction of a ~4 K radiation shield around the vacuum can containing the sapphire resonator, and better thermal grounding of the coaxial cables at the 4 K stage. Also, the same time domain data was converted to $S_y$ [dB/Hz] from the power spectral density of fractional frequency measurements after the linear drift was removed.  This is discussed below.
\begin{figure}[!t]
\centering
\includegraphics[width=2.5in]{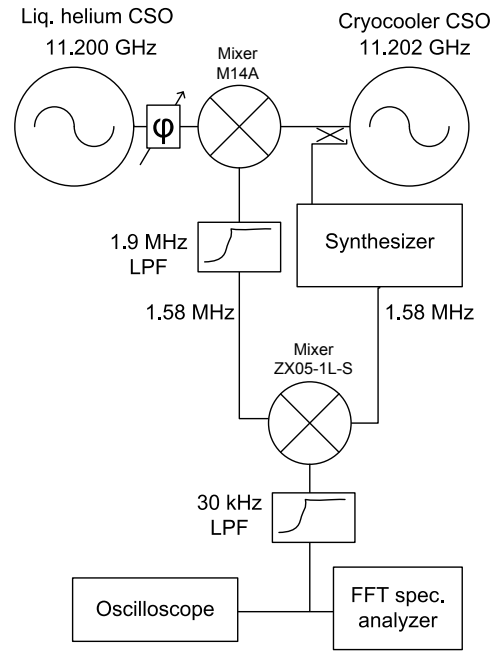}
\caption{Phase noise measurement setup used to obtain zero beat between the two cryogenic oscillators (see text for details).}
\label{fig_8}
\end{figure}

\section{Phase noise}
To date there have been very few phase noise measurements of cryogenic sapphire oscillators. Tobar et al. \cite{Tobar1994} measured the phase noise above 10 Hz Fourier frequencies in two nominally identical superconducting cavity oscillators, Dick et al. \cite{Dick1991} measured the phase noise of a superconducting sapphire cavity stabilized maser, Watabe et al. \cite{Watabe2007} measured the phase noise of two 100 MHz signals synthesized from two independent cryogenic sapphire oscillators, Marra et al. \cite{Marra}  measured the phase noise of two nominally identical liquid-helium cooled  sapphire oscillators and Grop et al. \cite{Grop2} measured the phase noise of a cryocooled  sapphire oscillator using a liquid-helium cooled oscillator. 

Recently we also estimated the low Fourier frequency phase noise from time domain measurements of the beat (sampled at 1 s)  between our cryocooled  sapphire oscillator and a liquid helium cooled sapphire oscillator in the same lab \cite{Hartnett2010}.  In the following we show measurement of the phase noise from those same two cryogenic oscillators but using the zero beat derived with the setup shown in Fig. \ref{fig_8}. 

\begin{figure}[!t]
\centering
\includegraphics[width=3.5in]{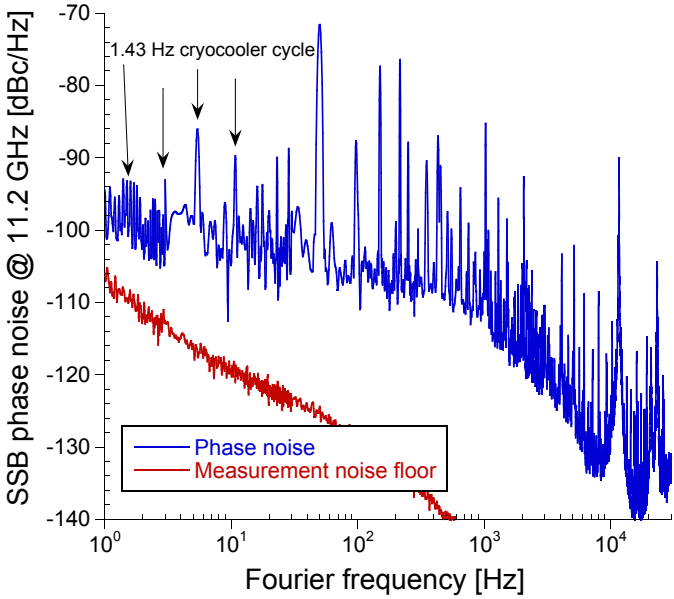}
\caption{(color online) The single side band phase noise $S_{\phi}$ [dBc/Hz] for a single oscillator assuming both contribute equally. }
\label{fig_9}
\end{figure}

Since the beat note of the two oscillators is about 1.58 MHz, a few mW of signal power at this frequency was derived from the IF port of a Watkins Johnson M14A mixer. This frequency was also synthesized from a low noise synthesizer that is phase locked to the cryocooled  sapphire oscillator. The details of this synthesizer can be found in Hartnett et al. \cite{Hartnett2009} (see Fig. 5 of that paper). The single side band residual phase noise of the down-converter in the synthesizer was measured to be -124 dBc/Hz at 1 Hz offset on a 100 MHz signal and hence does not contribute to the noise measurements here. See Fig. 3 of Hartnett et al. \cite{Hartnett2009}.  Using 1.58 MHz from the synthesizer as the LO drive on a Minicircuits ZX05-1L-S mixer we produced a zero beat. Final adjustment to get zero voltage on the oscilloscope was achieved with a mechanical microwave phase shifter in RF input arm to the M14A microwave mixer. Both cryogenic oscillators are sufficiently stable to maintain zero voltage, monitored on an oscilloscope, at the output IF port of the mixer, for at least a few minutes, sufficient to make the measurement on an Agilent 89410A FFT spectrum analyzer.  The resulting single sided phase noise for a single oscillator, assuming they both contribute equally,  is shown in Fig. \ref{fig_9}.  

The arrows indicate the bright lines derived from the cryocooler compressor cycle at 1.43 Hz. There is no clear bright line at the fundamental frequency of the cryocooler cycle, but it is clear at the higher harmonics. The hump at a few kHz is due to the bandwidth of the Pound servo used to lock the loop oscillator to the cryogenic resonator. Other lines are mains power harmonics. 

\begin{figure}[!t]
\centering
\includegraphics[width=3.5in]{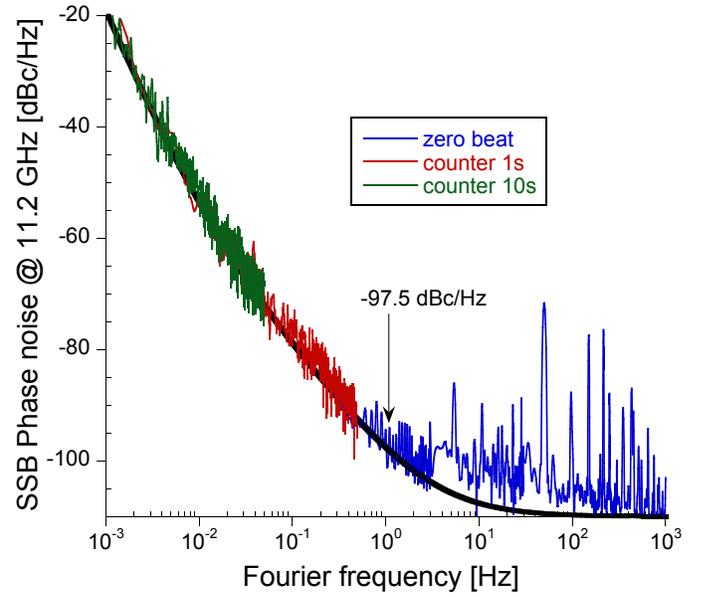}
\caption{(color online) The single side band phase noise $S_{\phi}$ [dBc/Hz] from $10^{-3}$ to $10^3$ Hz. The data labeled as counter 1 s and counter 10 s is the phase noise calculated from power spectral density of time domain beat data using a $\Lambda$-counter, with gate times of 1 s and 10 s, respectively.  The solid line is the best fit over the whole frequency range. }
\label{fig_10}
\end{figure}

In order to fully understand and characterize the different phase noise mechanisms observed in the oscillators we converted the power spectral density of time domain beat data to phase noise and show the result in Fig. \ref{fig_10} together with  phase noise data from the zero beat measurement (Fig. \ref{fig_9}). Using a $\Lambda$-counter \cite{Dawkins}, with gate times of 1 s and 10 s, the beat frequency was sampled and converted  to $S_y$ using the software \textit{Stable32}, after any linear drift was removed. 

Then the single side band phase noise was calculated  using,
\begin{equation} \label{Sy}
{\mathcal L}_{\phi}(f)=\frac{1}{2}\frac{\nu_0^2}{f^2} S_y(f),
\end{equation}
where $\nu_0$ is the microwave oscillator frequency. 

We then found the best fit power series that fits the measured power spectral density of phase fluctuations over each decade of offset Fourier frequencies in Fig. \ref{fig_10}. Hence the single oscillator single side band phase noise can be represented by 
\begin{eqnarray} \label{L}
& {\mathcal L}_{\phi}(f) = \frac{10^{-14.0}}{f^4} + \frac{10^{-11.6}}{f^3} + \frac{10^{-10.0}}{f^2} + ... \nonumber \\
& + \frac{10^{-10.2}}{f} + 10^{-11.0}\,\, [rad^2/Hz],
\end{eqnarray}
for Fourier frequencies $10^{-3}<f<10^3$ Hz. To represent Eq. (\ref{L}) on the plot in Fig. \ref{fig_10} take $10 \; log ({\mathcal L}_{\phi}(f))$ with units [dBc/Hz]. From Eq. (\ref{L}) we get $\cal{L}$$_{\phi}(1 \; Hz) \approx -97.5$ dBc/Hz, i.e. at 1 Hz offset from the 11.2 GHz carrier.

Then in turn if we calculate the Allan deviation from Eq. (\ref{L}) we get the result shown as the dashed curve in Fig. \ref{fig_6}. Here we evaluate the coefficients $h_{-2}, h_{-1}, h_0, h_1, h_2$ to $S_y$ determined using Eqs (\ref{Sy}) and (\ref{L}), for a $\Lambda$-counter (see the last column of Table I of Ref. \cite{Dawkins}). From there we calculated the total contribution to the Allan deviation and have compared that to the measured $\sigma_y$ in Fig. \ref{fig_6}. 

As would be expected the fit is very good (since we have mostly used time domain data) except  at integration times $\tau < 1$ s. This may be due to two causes: 1) it is difficult to account for all the bright lines at Fourier frequencies $f > 1$ Hz and 2) there is significant dead time ($\tau_d$) (by percentage of the sample duration) at both $\tau = 0.25$ and $0.5$ s. Flicker PM introduces an error of the order 0.43 $\tau_d/\tau$ for a high precision $\Lambda$-counter \cite{Dawkins}.

\section{Discussion}
The modeled $\sigma_y = \sigma_0 \times 10^{-15}$ for a single cryogenic oscillator can be characterized by
\begin{equation} \label{modelsy}
\sigma_0^2 = 0.072 + \frac{0.204}{\tau^3} + \frac{0.051}{\tau^2} + \frac{1.063}{\tau} + 
 0.00121 \; \tau.
\end{equation}
Therefore it follows that, at integration times $\tau \ll 1$ s the dominant noise process is white PM characterized by $$\sigma_y \approx \frac{4.52 \times 10^{-16}}{\tau^{3/2}}.$$ 
This may be reduced by the introduction of a lower phase noise amplifier. Over the range $1<\tau<10$ s white FM noise is dominant and $$\sigma_y \approx \frac{1.03 \times 10^{-15}}{\tau^{1/2}}.$$ 
Then $\sigma_y$ reaches a minimum value ($3.9 \times 10^{-16}$) where there is a limiting noise floor due to a flicker FM process. This can be characterized by $$\sigma_y \approx 2.69 \times 10^{-16}.$$ Assuming one could reduce the white FM noise one could conceivably reach this flicker floor. A flicker floor  as high as $2 \times 10^{-15}$ has been previously observed resulting from a noisy ferrite circulator used on the reflection port of the cryogenic resonator \cite{Chang2000}. The cause of the limiting flicker floor deserves further investigation in these oscillators.

At longer integration times there is random walk FM noise described by $$\sigma_y \approx 3.47 \times 10^{-17} \tau^{1/2}.$$  At $\tau = $1000 s this means $\sigma_y \approx 1.1 \times 10^{-15}$ and at $\tau = $10,000 s
$\sigma_y \approx 3.5 \times 10^{-15}$ but in practice temperature and pressure changes in the ambient environment have an effect. See Fig. \ref{fig_7}.

\section{Concluding Remarks}
In the first design of the cryostat \cite{Hartnett2010} we reported a significant linear frequency drift but this has now been reduced to less than $5 \times 10^{-14}$/day. We believe this largely resulted from thermal gradients, which we reduced by the installation of a 4 K radiation shield. But still there is room for improvement.   

The major advantage of the cryocooled oscillator is that it does not suffer from the need for regular and reliable liquid helium supplies in remote sites. The cryocooler head only requires  maintenance after about 3 to 5 years of continuous operation. Because this design incorporates a small volume of liquid helium (that is continually reliquified) if the power fails for short periods the resonator will remain cold. It takes 20 minutes after power fails for the pressure to build sufficient to vent helium gas through the relief valve. However the resonator will remain cold for several hours, which allows one to easily restart the unit after power is restored. If left too long more pure helium gas will need to be added. Nevertheless some hours of power failure can be accommodated by the design. 

The frequency stability of the cryocooled oscillator is sufficient to meet the requirements of a flywheel oscillator for atomic fountain clocks to overcome the Dick effect \cite{Wynands2005} and is more practical for standards labs than the version that required continual refilling with liquid helium. Its frequency stability matches that of the best liquid helium cooled cryogenic sapphire oscillators. It can be used as a local oscillator a few orders of magnitude more stable, at integration times between 1 and 10 s, than the best hydrogen maser, whereas at about 1000 s the stability becomes comparable to that of a hydrogen maser. 

The hydrogen maser is currently the standard in VLBI radio-astronomy, but at millimeter wave frequencies this can limit the coherence of the receiver signal. A more stable local reference can overcome this limitation. Therefore the cryocooled  sapphire oscillator has the potential to provide a much improved local oscillator for remote very high frequency VLBI radio-astronomy sites.

\section{Acknowledgments}
The authors would like to thank the Australian Research Council, Poseidon Scientific Instruments, the University of Western Australia, Curtin University of Technology and the CSIRO ATNF (Australian Telescope National Facility); the latter will provide a telescope site where the cryogenic sapphire oscillator will be tested as a local oscillator for VLBI radio astronomy. We also wish to thank E.N. Ivanov and M.E. Tobar for their useful advice, and J-M. Le Floch for assistance with data acquisition software.

%








\end{document}